\documentclass[11pt,twoside]{article}

\usepackage{asp2006}
\usepackage{epsf}
\usepackage{lscape}
\usepackage{graphicx}

\usepackage[OT4]{fontenc}                 

\begin{document}
\title{Four unusual novae observed in Toru\'n:\\V2362 Cyg, V2467 Cyg, V458 Vul, V2491 Cyg}   
\author{E. Ragan$^{1}$, M. Miko\l{}ajewski$^{1}$, T. Tomov$^{1}$, W. \rm Dimitrow$^{3}$, M. Fagas$^{3}$, 
T. Kwiatkowski$^{3}$, A. Schwarzenberg-Czerny$^{2,3}$, Ch. Buil$^{4}$, E.
Swierczy\'nski$^{1}$, 
T. Bro\.zek$^{1}$, M. Cika\l{}a$^{1}$, K. Czart$^{1}$, A. Fidos$^{1}$, S. Fr\k{a}ckowiak$^{1}$, C. Ga\l{}an$^{1}$, A. Karska$^{1}$,  M. K\l{}osi\'nska$^{1}$, M. Lewandowski$^{1}$, T. Radomski$^{1}$, P. R\'o{\.z}a\'nski$^{1}$, M. Wi\k{e}cek$^{1}$, P.
Wychudzki$^{1}$, A. Zajczyk$^{1}$, M. Zieli\'nska$^{1}$}   
\affil{$^1$Uniwersytet Miko\l{}aja Kopernika, Toru\'n, Polska\\
$^2$Centrum Astronomiczne Miko\l{}aja Kopernika, Warszawa, Polska\\
$^3$Uniwersytet Adama Mickiewicza, Pozna\'n, Polska\\
$^4$Castanet Tolosan Observatory, France}
\begin{abstract} 
We present photometric and spectral observation for four novae: V2362 Cyg, V2467 Cyg, V458 Vul, V2491 Cyg. All objects belongs to the ``fast novae'' class. For these stars we observed different departures from a typical behavior in the light curve and spectrum.
\end{abstract}
\section*{Introduction}
The objects were observed at Torun observatory as a part of the ,,Targets of opportunity'' project.
Spectroscopic and photometric observations were obtained between 2006 and 2008.
The photometric data were collected using the 60cm Cassegrain telescope and 60/90cm Schmidt Camera. The telescopes are equiped respectively with the CCD SBig STL-1001E and STL-11000M cameras. We also used the data from \textit{SAVS} project \citep{Nied03} and the AAVSO association \citep{Hen060708} .
Low resolution ($R\sim1500$) and a objective prism spectroscopy ($R\sim2500$ near H$\beta$) were taken with the Canadian Copernicus Spectrograph (CCS) and the flint prism attached to the 60/90cm Schmidt-Cassegrain telescope. Medium resolution ($R\sim6800$) echelle spectra were carried out with 28cm telescope in Castanet Tolosan observatory in France.
We also used high resolution echelle spectra ($R\sim35000$) obtained with a 40cm Newton telescope in Poznan.

\section*{V2362 Cyg}
Nova V2362 Cyg was discovered by H. Nishimura on Apr. 2, 2006 at $10^{\rm m}.5$
\citep{Nak06} and classified by \citet{Siv06} as a ,,Fe~II'' type. The light curve shows unusual brightening between 130 and 250 day after the maximum (Fig.~1). \citet{Kim08} shown that this event contained about $40\%$ of the total energy radiated during the whole outburst. They also found main properties: the maximum time Apr. 6, 2006, the brightness $m_{V}=7^{\rm m}.8$, the decline times $t_{2}=9^{\rm d}$, $t_{3}=21^{\rm d}$ and the distance $D\approx 7.5$ kpc. The spectra obtained near to both maxima show double P~Cyg absorptions in the emission line H$\alpha$ (Fig.~1). At the first maximum the expansion velocities were $-770$ and $-1750$~km~s$^{-1}$, between maxima the P~Cyg absorption disappeared and emerged in the second maximum with highest velocities $-1530$ and $-2280$~km~s$^{-1}$. Simultaneously the H$\alpha$ emission has been significantly broader at the second maximum. An additional red-shifted emission appeared, however the total flux in H$\alpha$, did not increase more than twice.
\begin{figure}[!ht]
\begin{center}
\includegraphics[scale = 0.25]{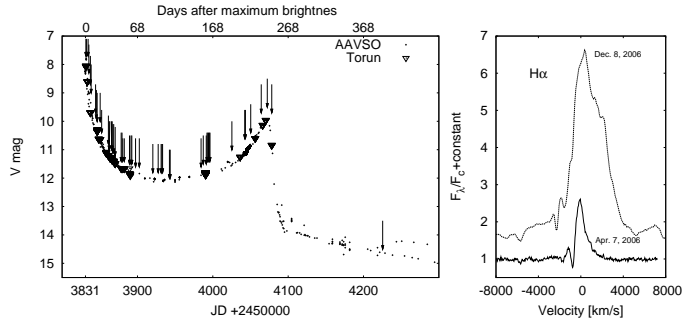}
\end{center}
\caption{Left: The light curve for V2362 Cyg. The open triangles indicate our photometry, the black diamonds -- AAVSO data and  the arrows -- the times of our spectral observations. Right: H$\alpha$ profiles around the maxima normalized to the local continuum.}
\end{figure}
\section*{V2467 Cyg}
Nova V2467~Cyg was discovered by Tago \citep{Nak07} on Mar. 15.8, 2007 at $V=7^{\rm m}.4$. The early spectrum obtained on Mar. 16.8, 2007 showed the expansion velocity of about $1200~{\rm km~s}^{-1}$ in the P~Cyg absorption in H$\alpha$ emission line \citep{Nak07}. This object was classified by \citet{Mun07} as ``FeII'' type. 
\citet{Ste07} estimated the distance between 1.5--4 kpc and the outburst amplitude $\sim 12^{\rm m}$. Our light curve of V2467~Cyg is shown in Fig.~2. The spectra of V2467~Cyg between Apr. 13, 2007 and May 16, 2007 were dominated by the Balmer emission lines and the extremely strong OI 8446\AA\ emission line, which flux was grater than the flux observed in H$\alpha$ line. The photon emitted by OI 8446\AA\ can be strengthened as a result of pumping of the Ly$\beta$ photons \citep{Bow47}, and the observed fluxes suggest an overabundance of oxygen in this object \citep{Tom07} .
\begin{figure}[!ht]
\begin{center}
\includegraphics[scale = 0.25]{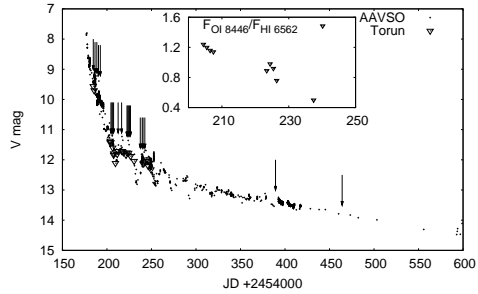}
\end{center}
\caption{The light curve of V2467 Cyg. The open triangles indicate our photometry, the black diamonds -- AAVSO data and arrows -- the times of our spectral observations. The inserted box shows the changes of OI 8446\AA\ to H$\alpha$ flux ratio.}
\end{figure}
\section*{V458 Vul}
Nova Vulpeculae 2007 was discovered by H. Abe \citep{Nak07a} on 2007 Aug 8.54 UT. at $V=8^{\rm m}$. In the beginning the object was classified as a ``Fe~II'' type nova and later as a hybrid nova \citep{Tar07} . The nova presented an unusual light curve (Fig.~3) with fast fading and three short maxima separated by 3--5 days. At each of the maximum we measured velocities of P~Cyg absorption observed in H$\alpha$ emission line which were slightly increasing in following maxima. In Fig.~3 we show the evolution of HeI 6678\AA. Most characteristic in this line seems to be the ,,saddle shape'' profiles visible after each of the maxima. The full width at zero intesity (FWZI) of this line measured one day after the first, second and the third maximum, was $\sim 3000$, $\sim 4400$, $\sim 5000$ km~s$^{-1}$ respectively. The most probably, we can see the three independent ejections with the increasing velocities.
\begin{figure}[!ht]
\begin{center}
\includegraphics[scale = 0.2]{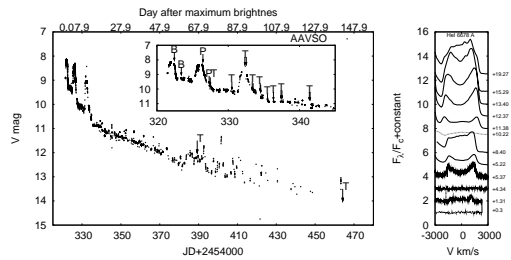}
\end{center}
\caption{Left: the light curve for V458 Vul. The black diamonds indicate the AAVSO photometric data, the arrows -- the times of spectral observations from Toru\'n (T), Pozna\'n (P) and Castanet Tolosan (B) observatories. Right: the evolution of the HeI 6678\AA\, emission line. The thin lines indicate the spectra obtained in maxima (when the profiles were almost flat), the thick lines -- between them.}
\end{figure}
\section*{V2491 Cyg}
Nova V2491 Cyg was detected by F. Kabashima and K. Nishiyama on Apr. 10.7, 2008 at $7^{\rm m}.7$ \citep{Nak08} . The very broad H$\alpha$ emissions with FWHM of $\sim 4000$ km~s$^{-1}$ were visible on our first spectra on Apr. 11.99, 2008 and Apr. 13.95 \citep{Tom08a}. We classified this nova as a ,,He/N'' type with a post-maximum spectrum of subclass $P_n^o$ \citep[see][]{Wil91}, and also suggested its similarity to spectra of the recurrent nova U Sco and V394 CrA. Moreover, V2491 Cyg is the second nova which has been observed in X-rays before the outburst \citep{Ibr08}, the previous was V2487 Oph. On our early spectra, except Balmer emission lines, there were strong lines in the near infrared region: HeI 10083\AA, NI 8692, 8212\AA, MgII 9226\AA. Spectra obtained between Apr. 15 and 25, 2008 showed the double P~Cyg absorptions in Balmer emission lines with the velocities from $\sim 2700$ to $\sim 4300$ km~s$^{-1}$ for the slower, and from $\sim 3800$ to 6400 km~s$^{-1}$ for the faster component. The light curve, the color variations and the evolution of H$\alpha$ and OI 8446\AA\ profiles are shown in Fig.~4.
\begin{figure}[!ht]
\begin{center}
\includegraphics[scale = 0.19]{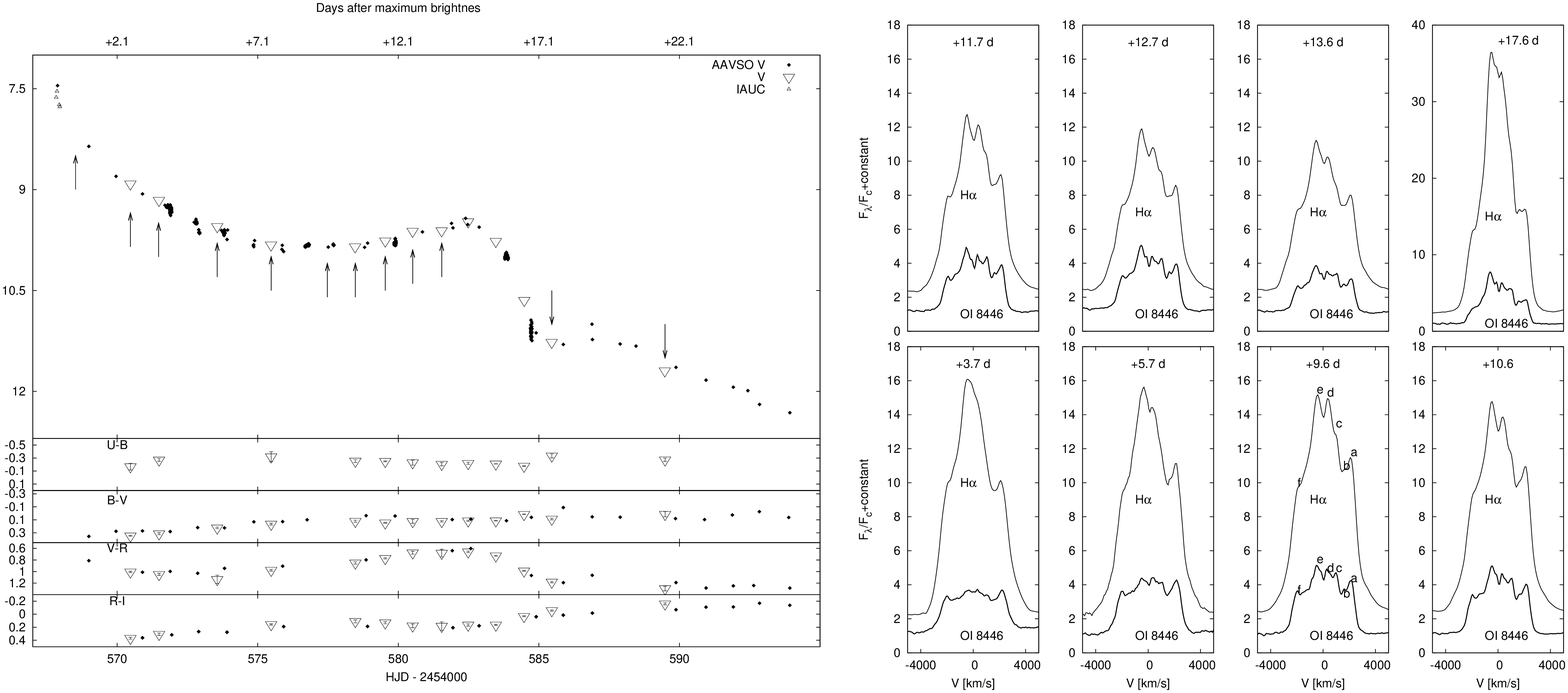}
\end{center}
\caption{\small Left: The light curve and the color variations of V2491 Cyg. The open triangles indicate our photometry, the black diamonds -- AAVSO data, the black triangles -- IAUC publication \citep{Nak08} and the arrows -- the times of our spectral observations. Right: evolution of the OI 8446\AA\ and the H$\alpha$ emission line.}
\end{figure}
\acknowledgements 
We acknowledge with thanks the variable star observations from the AAVSO International Database contributed by observers worldwide and used in this research. This work was supported by the Polish MNiSW Grant N203 018 32/2338 and MNiI/MNiSzW grant 1P03D 025 29.



\end{document}